\def\beqa{\begin{eqnarray}}
\def\eeqa{\end{eqnarray}}

\def\ra{\rangle}
\def\la{\langle}
\def\be{\begin{equation}}
\def\ee{\end{equation}}
\def\ba{\begin{array}}
\def\ea{\end{array}}

\def\ra{\rangle}
\def\la{\langle}

\def\qed{\leavevmode\unskip\penalty9999 \hbox{}\nobreak\hfill
     \quad\hbox{\leavevmode  \hbox to.77778em{%
               \hfil\vrule   \vbox to.675em%
               {\hrule width.6em\vfil\hrule}\vrule\hfil}}
     \par\vskip3pt}

\documentclass[preprint,aps,pre,amsmath,amssymb,amsfonts,showpacs]{revtex4}
\usepackage{epsfig}
\usepackage{amsmath}

\begin{document}
\title{Lower Bound of Concurrence Based on Positive Maps}
\author{Xiao-Sheng  Li$^{1}$}
\author{Xiu-Hong Gao$^{2}$}
\author{Shao-Ming Fei$^{2,3}$}
\affiliation{$^1$Department of Mathematics, School of Science, South
China University of Technology, Guangzhou 510640, China\\
$^2$School of Mathematical Sciences, Capital Normal
University, Beijing 100048, China\\
$^3$Max-Planck-Institute for Mathematics in the Sciences, 04103
Leipzig, Germany}

\begin{abstract}
We study the concurrence of arbitrary dimensional bipartite quantum
systems. An explicit analytical lower bound of concurrence is
obtained, which detects entanglement for some quantum states better
than some well-known separability criteria, and improves the lower
bounds such as from the PPT, realignment criteria and the Breuer's
entanglement witness.
\end{abstract}

 \maketitle

Quantum entangled states are the most important resource in quantum
information processing \cite{hhhh}. An important theoretical
challenge in the theory of quantum entanglement is to give a proper
description and quantification of quantum entanglement for given
quantum states. The entanglement of formation (EOF) \cite{eof} and
concurrence \cite{con} are two well defined quantitative measures of
quantum entanglement. For the two-qubit case EOF is a monotonically
increasing function of the concurrence and an elegant formula for
the concurrence was derived analytically by Wootters in
\cite{Wootters98}. It plays an essential role in describing quantum
phase transitions in various interacting quantum many-body systems
\cite{Osterloh02-Wu04,Ghosh2003} and can be experimentally measured
\cite{buchleitner}. However for general high dimensional case, due
to the extremizations involved in the calculation, only a few
explicit analytic formulae for EOF and concurrence have been found
for some special symmetric states \cite{Terhal-Voll2000}.

In stead of analytic formulae, some progresses have been made toward
the lower bounds of EOF and concurrence. For instance,
in \cite{167902} a lower bound of concurrence
that can be tightened by numerical optimization over some parameters
has been derived. In \cite{Chen-Albeverio-Fei1,chen} analytic
lower bounds on EOF and concurrence for any dimensional mixed
bipartite quantum states have been presented by using the positive partial
transposition (PPT) and realignment separability criteria.
These bounds are exact for some special classes of states and can be used to detect many bound
entangled states. In \cite{breuer} another lower bound on EOF for even dimensional
bipartite states has been presented from a new separability
criterion \cite{breuerprl}. A lower bound of concurrence based on
local uncertainty relations (LURs) criterion is derived in
\cite{vicente}. This bound is further optimized in \cite{zhang}.
In \cite{edward,ou} the authors presented
lower bounds of concurrence for bipartite systems in terms of
a different approach, which has a
close relationship with the distillability of bipartite quantum states.
In \cite{limingb} an experimentally measurable bounds for EOF has been presented.
All these bounds obtained so far together give rise to a good quantitative estimation of
EOF and concurrence. In particular, they are supplementary in detecting
entanglement.

In this brief report, based on positive maps, we present a new lower bound of concurrence for
arbitrary dimensional bipartite systems. This bound is shown to detect
entanglement that can not be recognized by the bounds in
\cite{chen,breuer}.

Let $H_1$ and $H_2$ be $N$-dimensional vector spaces.
A bipartite quantum pure state $|\psi\ra$ in $H_1\otimes H_2$ has a
Schmidt form
\begin{equation}
\label{schmit}
|\psi\ra=\sum_i \alpha_i|e_i^1\ra \otimes|e_i^2\ra,
\end{equation}
where $|e_i^1\ra$ and  $|e_i^2\ra$  are the orthonormal bases in $H_1$ and
$H_2$ respectively, $\alpha_i$ are the Schmidt coefficients satisfying $\sum_i\alpha_i^2=1$.

The concurrence of the state $|\psi\ra$ is given by
\be\label{concurrence}
C(|\psi\ra)=\sqrt{2(1-Tr\rho_1^2})=2\sqrt{\sum_{i<
j}\alpha_i^2\alpha_j^2}, \ee where the reduced density matrix
$\rho_1$ is obtained by tracing over the second subsystem of the
density matrix $\rho=|\psi\ra\la \psi|$, $\rho_1=Tr_2|\psi\ra\la
\psi|$.

A general mixed state in  $H_1\otimes H_2$ has pure state
decompositions, $\rho=\sum_i p_i|\psi_i\ra \la\psi_i|$, where $p_i
\geq 0$ and $\sum_i\ p_i=1$. The concurrence is extended to mixed
states $\rho$ by the convex roof, \be\label{concurrencem}
C(\rho)=min \sum_i\ p_i C(|\psi_i\ra), \ee where the minimum is
taken over all possible convex decompositions of $\rho$ into an
ensemble $\{|\psi_i\ra\}$ of pure states with probability
distribution $\{p_i\}$.

Let $f(\rho)$ be a real-valued and convex functional on the
total state space with the following property,
\begin{equation} \label{property}
f(|\psi\ra \la\psi|)\leq 2\sum_{i< j}\alpha_i\alpha_j,
\end{equation}
for all state vectors $|\psi\ra$ with Schmidt decompositions (\ref{schmit}).
By using the inequality in \cite{chen},
$$
\sum_{i<j}\alpha_{i}^2\alpha_{j}^2 \geq \frac{2}{N(N-1)}\left(\sum_{i<j}\alpha_{i}\alpha_{j}\right)^{2},
$$
Breuer \cite{breuer} has derived that $C(\rho)$ satisfies
\be\label{bb} C(\rho)\geq\sqrt{\frac{2}{N(N-1)}} f(\rho). \ee
The $f(\rho)$ corresponding to the lower bounds in \cite{chen} are the ones
with respect to the PPT criterion and the realignment criterion,
$f_{ppt}(\rho) = ||\rho^{T_1}||-1$, $f_{r}(\rho) =
||\tilde{\rho}||-1$, where $||\cdot||$ stands for the trace norm of
a matrix, $T_1$ the partial transposition associated with the space
$H_1$ and $\tilde{\rho}$ the realigned matrix of $\rho$. While the
lower bound in \cite{breuer} corresponds to a convex functional
$f_W(\rho) = -Tr(W\rho)$, where $W$ is the entanglement witness
introduced in \cite{breuer}.

Let $\Phi$ be a matrix map that maps an $N\times N$ matrix $A$, $(A)_{ij}=a_{ij}$, $i,j=1,...N$,
to an $N\times N$ matrix $\Phi(A)$ with $(\Phi(A))_{ij}=-a_{ij}$ for $i\neq j$, and
$(\Phi(A))_{ii}=(N-2)a_{ii}+a_{i^\prime i^\prime}$, where $i^\prime=i+1~mod~ N$.
It can be shown that the matrix map $\Phi$ is positive but not completely positive \cite{Hou}.

{\bf Theorem} \  For any bipartite quantum state $\rho=\sum_i p_i|\psi_i\ra\la\psi_i|\in H_1\otimes H_2$,
the concurrence $C(\rho)$ satisfies
\begin{equation}\label{theorem}
C(\rho)\geq \sqrt{\frac{2}{N(N-1)}} (\|(I_N\otimes\Phi)\rho\|-(N-1)),
\end{equation}
where $I_N$ is the $N\times N$ identity matrix.

{\bf Proof} \ Set $f(|\psi\ra\la\psi|)=\|(I_N\otimes\Phi)|\psi\ra\la\psi|\|-(N-1)$.
Obviously $f(|\psi\ra\la\psi|)$ is convex as the trace norm is convex. What we need to prove is that,
for any pure state in Schmidt form (\ref{schmit}), the inequality (\ref{property}) holds.

Since the trace norm does change under local coordinate transformation,
we take $|\psi\ra=(\alpha_{1},0,\cdots,0,0,\alpha_{2},
\cdots,0,0,0,\alpha_{3},\cdots,0,\cdots\cdots,0,\cdots,0,\alpha_{N})^t$, where $t$
denotes transposition, the Schmidt coefficients satisfy $0\leq \alpha_{1}, \alpha_{2},
\alpha_{3},\cdots, \alpha_{N}\leq 1$,
$\alpha_{1}^2+\alpha_{2}^2+\alpha_{3}^2+\cdots+\alpha_{N}^2=1$.

It is direct to verify that the matrix $T\equiv (I_N\otimes \Phi )(|\psi\ra\la\psi|)$ has
$N^{2}-2N$ singular values $0$, $N$ singular values $\alpha_{1}^2, \alpha_{2}^2, \alpha_{3}^2,\cdots,
\alpha_{N}^2$, the remaining $N$ ones are the singular values of the following matrix $B$:
$$
\ba{l}
B=
{\small \left(
\begin{array}{ccccc}
(N-2)\alpha_{1}^2&-\alpha_{1}\alpha_{2}&-\alpha_{1}\alpha_{3}&\cdots &-\alpha_{1}\alpha_{N}\\[1mm]
-\alpha_{1}\alpha_{2}&(N-2)\alpha_{2}^2&-\alpha_{2}\alpha_{3}&\cdots &-\alpha_{2}\alpha_{N}\\[1mm]
\vdots&\vdots&\vdots&\cdots&\vdots\\[1mm]
-\alpha_{1}\alpha_{N}&-\alpha_{2}\alpha_{N}&-\alpha_{3}\alpha_{N}&\cdots&(N-2)\alpha_{N}^2
\end{array}
\right). }
\ea
$$
As $B$ is Hermitian and real, its singular
values are simply given by the square roots of the eigenvalues of
$B^2$. In fact we only need to consider the absolute values of the eigenvalues of $B$.
The eigenpolynomial equation of $B$ is:
\begin{eqnarray} \label{root-coeff11}
H(x)&=&\mid xI_N-B \mid =x^N-(N-2)x^{N-1}+(N-3)(N-1)(\sum_{i<
j}\alpha_{i}^2\alpha_{j}^2)
x^{N-2}\nonumber\\
&&-(N-4)(N-1)^2(\sum_{i< j<
k}\alpha_{i}^2\alpha_{j}^2\alpha_{k}^2)x^{N-3}
+\cdots\nonumber\\
&&+(-1)^{N}(N-1)^{N-3}(\sum_{i_{1}< i_{2}\cdots<
i_{N-2}}\alpha_{i_{1}}^2\alpha_{i_{2}}^2\cdots\alpha_{i_{N-2}}^2)
x^2
\nonumber\\&&+(-1)^{N+1}(N-1)^{N-1}(\alpha_{1}^2\alpha_{2}^2\alpha_{3}^2\cdots\alpha_{N}^2)=0.
\end{eqnarray}

Let $x_1,x_2, x_3,\cdots, x_N$ denote the $N$ roots of Eq.(\ref{root-coeff11}).
By using the relations between roots and coefficients of the polynomial equation, one has
\begin{equation} \label{root-coeff44}
\sum_{i=1}^N x_i=N-2,~~~
\Pi_{i=1}^N x_i=(-1)^{2N+1}(N-1)^{N-1}\Pi_{i=1}^N \alpha_{i}^2.
\end{equation}
The inequality (\ref{property}) that needs to be proved has the form
now, \be\label{iqp} \sum_{i=1}^N|x_i|-(N-2)\leq  2(\sum_{i<
j}\alpha_{i}\alpha_{j}), \ee where $\sum_{i=1}^N \alpha_{i}^2 =1$
has been taken into account.

To deal with the eigenpolynomial equation (\ref{root-coeff11}), we set
$\beta=\Pi_{i=1}^N \alpha_{i}^2$.

(a) If $\beta=0$, then $H(0)=0$, $0$ is an eigenvalue of $B$. From the derivative of $H(x)$ with respect to
$x$,
\begin{eqnarray}
H'(x)&=&Nx^{N-1}-(N-2)(N-1)x^{N-2}+(N-3)(N-2)(N-1)(\sum_{i<
j}\alpha_{i}^2\alpha_{j}^2)x^{N-3}\\\nonumber
&&-\cdots +2(-1)^{N}(N-1)^{N-3}(\sum_{i_{1}< i_{2}\cdots<
i_{N-2}}\alpha_{i_{1}}^2\alpha_{i_{2}}^2\cdots\alpha_{i_{N-2}}^2)x,
\end{eqnarray}
we know that if $N$ is even, $H'(x)<0$ when $x<0$. Therefore $H(x)$ is a
monotonically decreasing function when $x<0$. Taking into account that $H(0)=0$, we see that
there exist no negative roots of (\ref{root-coeff11}) in this case.

The inequality (\ref{iqp}) that needs to be proved has the form now,
\begin{equation}\label{ine44}
\sum_{i=1}^N x_i -(N-2)\leq  2(\sum_{i< j}\alpha_{i}\alpha_{j}).
\end{equation}
According to the relation in (\ref{root-coeff44}), the left hand of the
inequality (\ref{ine44}) is zero. Hence the inequality (\ref{iqp}) is satisfied.

When $N$ is odd, $H(x)$ is a monotonically increasing function for $x<0$.
There are also no negative roots of (\ref{root-coeff11}). One can similarly prove the
inequality (\ref{iqp}).

(b) When $\beta\neq 0$, we have $H(0)=(-1)^{N+1}(N-1)^{N-1}(\alpha_{1}^2\alpha_{2}^2\alpha_{3}^2\cdots\alpha_{N}^2)$.
If $N$ is even, we have $H(0)<0$. From (\ref{root-coeff44}) we get $x_1 x_2 x_3\cdots x_N<0$. Therefore, there exists
at least one negative root, say $x_{1}<0$, such that $H(x_{1})=0$.

Due to that $H'(x)<0$ for $x<0$, $H(x)$  is a monotonically decreasing function when
$x<0$. Taking into account that $H(0)<0$, we have that $x_{1}<0$ is the only negative root.
Hence the inequality (\ref{iqp}) needed to be proved becomes:
\begin{equation}\label{ine44p}
\sum_{i=2}^N x_i-x_1-(N-2)\leq  2(\sum_{i< j}\alpha_{i}\alpha_{j}).
\end{equation}

From Eq. (\ref{root-coeff44}), we only  need to prove that $x_1\geq
-\sum_{i< j}\alpha_{i}\alpha_{j}$. From the definition of $H(x)$, we
have $H(-\sum_{i< j}\alpha_{i}\alpha_{j})=|-(\sum_{i<
j}\alpha_{i}\alpha_{j})I_N-B| =|(\sum_{i<
j}\alpha_{i}\alpha_{j})I_N+B| \geq 0$, where in the last step the
property of the diagonally dominant matrix $(\sum_{i<
j}\alpha_{i}\alpha_{j})I_{N}+B$ has been used. Since $H(x_1)=0\leq
H(-\sum_{i< j}\alpha_{i}\alpha_{j})$ and $H(x)$ is a monotonically
decreasing function when $x<0$, we have that $x_1\geq -\sum_{i<
j}\alpha_{i}\alpha_{j}$.

When $N$ is odd, $H(x)$ is a monotonically increasing function when $x<0$.
The theorem can be similarly proved. \qed

Our bound (\ref{theorem}) can detect better entanglement than other
lower bounds of concurrence. As an example let us consider a state
in $4\times 4$ \cite{Hou},
\begin{equation}\label{example}
\rho=(1/4) diag(q_1, q_4, q_3, q_2, q_2, q_1, q_4,
q_3, q_3, q_2, q_1, q_4, q_4, q_3, q_2,
 q_1)+\frac{q_1}{4} \sum_{i,j=1, 6, 11, 16}^{i\neq j} F_{i,j},
\end{equation}
where $F_{i,j}$ is the unit matrix with $(i,j)$-entry $1$ and others
$0$, $q_m\geq 0$, $\sum q_m=1$, $m=1, 2, 3, 4$.

For $N=4$, the positive map $\Phi$ maps a matrix $M$ with $(M)_{ij}=(m_{ij})$, $i, j=1,..., 4$, to
$$\ba{l} \Phi(M)=\left(
\begin{array}{cccc}
2m_{11}+m_{22}&-m_{12}&-m_{13}&-m_{14}\\[1mm]
-m_{21}&2m_{22}+m_{33}&-m_{23}&-m_{24}\\[1mm]
-m_{31}&-m_{32}&2m_{33}+m_{44}&-m_{34}\\[1mm]
-m_{41}&-m_{42}&-m_{43}&2m_{44}+m_{11}
\end{array}
\right).
\ea
$$
By direct computation we have the following set of eigenvalues of
$(I_4\otimes\Phi)(\rho)$:
\begin{eqnarray} \frac{1}{4}&\{q_1+2q_2,q_1+2q_2,q_1+2q_2,q_1+2q_2,q_2+2q_3,q_2+2q_3,
q_2+2q_3,q_2+2q_3, \nonumber\\&
q_3+2q_4,q_3+2q_4,q_3+2q_4,q_3+2q_4,q_4-q_1,3q_1+q_4,3q_1+q_4,3q_1+q_4\}.\nonumber
\end{eqnarray}
Therefore from (\ref{theorem}) we have
\begin{equation} \label{new bound}
C(\rho)\geq \sqrt{\frac{1}{6}}(\|(I_4\otimes \Phi)\rho\|-3)=\frac{1}{4\sqrt{6}}(q_1-q_4+|q_1-q_4|).
\end{equation}
From \cite{chen}, with respect to the PPT and realignment operation one has bounds
\begin{equation} \label{PPT bound}\ba{rcl}
C_{PPT}(\rho)&\geq& \sqrt{\frac{1}{6}}(\|\rho^{T_1}\|-1)\\
&=&\frac{1}{2\sqrt{6}}(2q_1+|q_1-q_3|+\left|q_2+q_4-\sqrt{4q_1^2+(q_2-q_4)^2}\right|\\[2mm]
&&+\sqrt{4q_1^2+(q_2-q_4)^2}-1)
\ea
\end{equation}
and
\begin{equation} \label{realignment bound}
\ba{rcl}
C_{r}(\rho)&\geq&\sqrt{\frac{1}{6}}(\|\tilde{\rho}\|-1)\\[2mm]
&=&\sqrt{\frac{1}{6}}\left(3q_1+\frac{1}{4}(\sqrt{(q_1-q_2+q_3-q_4)^2}+2\sqrt{(q_1-q_3)^2+(q_2-q_4)^2}-3)\right).
\ea
\end{equation}
From the formula presented in \cite{breuer}, one has the bound
\begin{equation} \label{breuer bound}
C_W(\rho)\geq \sqrt{\frac{1}{6}}(-tr(W\rho))=-\frac{1}{2\sqrt{6}}(q_2+2q_3+q_4).
\end{equation}

\begin{figure}[htb]
\begin{center}
\includegraphics{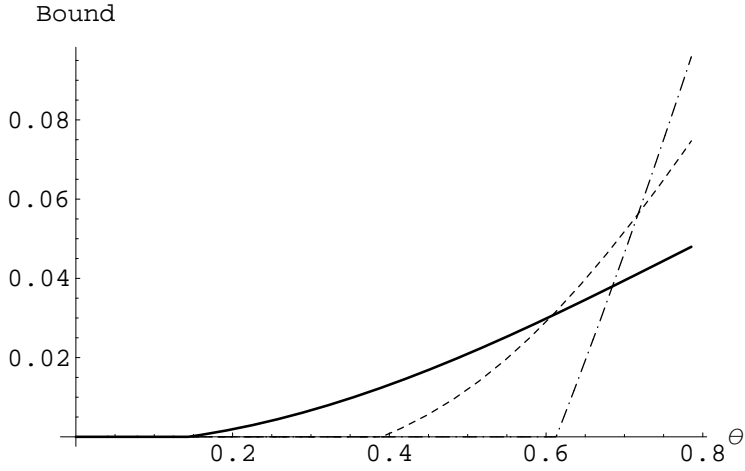}
\end{center}
\caption{Lower bounds for the state (\ref{example}). Solid line: the
lower bound given by (\ref{new bound}); Dashed line: lower
bound given by (\ref{PPT bound});
Dashed-dotted line: lower bound given by (\ref{realignment bound});
$\theta$ axis: lower bound given by (\ref{breuer bound}).}
\end{figure}

To compare among these bounds, let us take $q_2=\frac{1}{2}$,
$q_4=0.01$, $q_1=(1-q_2-q_4)sin^2\theta$,
$q_3=(1-q_2-q_4)cos^2\theta$, $\theta\in [0, \frac{\pi}{4}]$. From
Fig. 1 we see that our new bound (\ref{new bound}) detects
entanglement for $\theta>0.143$. While $C_{PPT}$ and $C_{r}$ detect
entanglement for $\theta>0.390 $ and $\theta>0.613$ respectively, and $C_W$ can
not detect any entanglement as it is always negative.

In summary, by using a positive map we have presented a new lower
bound of concurrence for arbitrary dimensional bipartite systems. By
a detailed example we have shown that this bound is better for some
states than the lower bounds from the PPT criterion, the realignment
criterion and the Breuer's entanglement witness \cite{breuer} in
detecting quantum entanglement.

\bigskip
\noindent{\bf Acknowledgments}\, This work was supported by the NSFC
(10875081,10801100,10871228), KZ200810028013 and PHR201007107.

\end{document}